\documentclass[conference]{IEEEtran}
\IEEEoverridecommandlockouts
\usepackage{cite}
\usepackage{amsmath,amssymb,amsfonts}
\usepackage{algorithmic}
\usepackage{graphicx}
\usepackage{subcaption}
\usepackage{textcomp}
\usepackage{booktabs}
\usepackage{xcolor}
\usepackage{comment}
\usepackage{acronym}
\usepackage{tabularx}
\def\BibTeX{{\rm B\kern-.05em{\sc i\kern-.025em b}\kern-.08em
    T\kern-.1667em\lower.7ex\hbox{E}\kern-.125emX}}
\begin{document}

\newacro{gur}[GUR]{Games User Research}
\newacro{sna}[SNA]{Social Network Analysis}
\newacro{osns}[OSNs]{Online Social Networks}
\newacro{f2p}[F2P]{Free-to-play}

\definecolor{ballblue}{rgb}{0.13, 0.67, 0.8}
\newcommand{\todo}[1]{\marginpar{\color{ballblue}{\texttt{TODO}}} {\color{ballblue}{#1}}}
\newcommand{\checkcontent}[1]{\marginpar{\color{orange}{\texttt{CHECK}}} {\color{orange}{#1}}}

%8pg ref included
\title{Comparing the Structures and Characteristics of Different Game Social Networks - The Steam Case}
%\title{Let's Be Friends -- How Games' Characteristics Reflect on the Steam Friends Network Structure}

\author{\IEEEauthorblockN{Enrica Loria}
\IEEEauthorblockA{\textit{ISDS Institute} \\ \textit{Graz University of Technology}\\
Graz, Austria \\
eloria@tugraz.at}
\and
\IEEEauthorblockN{Alessia Antelmi}
\IEEEauthorblockA{\textit{Dipartimento di Informatica} \\
\textit{Università degli Studi di Salerno}\\
Salerno, Italy \\
aantelmi@unisa.it}
\and
\IEEEauthorblockN{Johanna Pirker}
\IEEEauthorblockA{\textit{ISDS Institute} \\ \textit{Graz University of Technology}\\
Graz, Austria \\
johanna.pirker@tugraz.at}
}

\IEEEpubid{\begin{minipage}{\textwidth}\ \\[12pt]
978-1-6654-3886-5/21/\$31.00 \copyright 2021 IEEE
\end{minipage}}

\maketitle

\begin{abstract}
In most games, social connections are an essential part of the gaming experience. Players connect in communities inside or around games and form friendships, which can be translated into other games or even in the real world. Recent research has investigated social phenomena within the player social network of several multiplayer games, yet we still know very little about how these networks are shaped and formed. Specifically, we are unaware of how the game type and its mechanics are related to its community structure and how those structures vary in different games. This paper presents an initial analysis of Steam users and how friendships on Steam are formed around 200 games. We examine the friendship graphs of these 200 games by dividing them into clusters to compare their network properties and their specific characteristics (e.g., genre, game elements, and mechanics). We found how the Steam user-defined tags better characterized the clusters than the game genre, suggesting that how players perceive and use the game also reflects how they connect in the community. Moreover, team-based games are associated with more cohesive and clustered networks than games with a stronger single-player focus, supporting the idea that playing together in teams more likely produces social capital (i.e., Steam friendships).

\end{abstract}


\section{Introduction}
%\begin{itemize}
%    \item Wide context
%    \item Knowledge Gap/Motivation
%    \item Contribution/So what?
%\end{itemize}
Well-being~\cite{Johannes2020VideoWell-being}, increased engagement~\cite{Przybylski2010AEngagement,Pirker2018AnalyzingDestiny}, longer-term retention~\cite{Schiller2018InsideActivity}, and a sense of connectedness~\cite{Mandryk2020HowWellbeing} are established benefits deriving from playing together. Researchers increasingly show that playing video games is a social activity~\cite{Nacke2015SocialGame,Johannes2020VideoWell-being,Mandryk2020HowWellbeing}. Players gather in online and offline communities revolving around games, not only relying on the embedded social mechanics of the game but also on social media~\cite{wallner2019tweeting} and other forums~\cite{Wallner2019BeyondWebsite}. This enhances the social aspect of single-player games, for which a community of affectionate players can also exist and connect. Studying the player community provides much information on how the game is perceived and also on the gameplay experience. Consequently, the analysis of the player social network (either implicit and explicit) is a practice that is catching on among researchers and game analysts. Most of the research on \ac{sna} in games is in the realm of strongly multiplayer games (often team-based), such as League of Legends~\cite{Kim2017WhatLegends}, World of Warcraft~\cite{Mandryk2020HowWellbeing}, and Destiny~\cite{Pirker2018AnalyzingDestiny,loria2020influencers}, hence neglecting single-player games communities. Those works are generally interested in connecting social interaction patterns to players' activity and engagement in the specific context without analyzing the community at a higher level and making a cross-game comparison. Players connect in the game, social platforms, and game providers. As a result, the investigation of the social nature of games is more complex than just analyzing in-game social networks. Nevertheless, we are still unaware of how and whether the characteristics of games reflect on the way players connect and which are the design elements or mechanics fostering more cohesive communities. Researchers argue that games build friendships, which have value in the real world and are transferred across games. However, a still unanswered question is: how does the type of game impact the formation of those friendships?

We analyzed a game provider, Steam\footnote{https://steampowered.com} as a step towards the analysis of player friendships. The Steam platform is a great incubator of social relationships among players due to its hybrid nature combining functionality and few social media elements. Steam is primarily a game library where users can buy and store their games, build a profile, and share their activities and achievements. Additionally, Steam allows players to institute explicit friendship links with each other and obtain updates on the actions of friends (e.g., playing, reviewing, and buying new games) and progress. Specifically, we collected and analyzed parts of the Steam friends network and filtered it by each game played from the population sample. We then clustered the game friends sub-graph in six clusters. We characterized the clusters according to their genre and user-defined tags and the network properties (e.g., clustering coefficient, degree distribution, and centralization).

\subsection*{Research Goal and Contributions}
In summary, the \ac{gur} literature provides evidence on the importance of sociality in games and the community built around games. Yet, the existing investigations into the social network of the players are conducted on specific games and with clear objectives in mind (e.g., studying retention, social influence, and engagement). Consequently, we are unaware of the social network characteristics that can be used to describe different game types. 
This study aims to perform an exploratory investigation to identify how the game characteristics, expressed by Steam genres and user-defined tags, relate to the shape player community. Our results connect some tags to specific network properties, contributing to our understanding of how users conceive the games (i.e., tag them) reflects on the specific game community structure. For instance, single-player games have scattered networks, whereas multiplayer games show more connected graphs, especially when team-based. Those networks are often scale-free and thus resemble the degree distribution of social media networks. Unlike tags, the game genres cannot be related to community characteristics. 
As a result, our contribution to the \ac{gur} community is twofold. First, we show how the definition of games through tags by the players (in the Steam platform) can be more meaningful for the understanding of the community than the genre originally associated with the game by Steam (or the designers). In other words, the way the game is perceived and used by the players provides a good picture of the actual player network shape. On the other hand, the game genre does not provide additional information. Second, we show how including a team-based (and often local) multiplayer mechanics in games would more likely produce a cohesive, connected community of friends that resembles (in structure) social media networks. This finding further supports the conception of multiplayer (team-based) games as social incubators.

\section{Related Work}
%\todo{update/add more recent refs\\}
%\subsection*{Social Games}
People seek relatedness in any kind of activity~\cite{Baumeister1995TheMotivation}, and very frequently do so in games~\cite{Rogers2017TheApproach}. Games embed a degree of social connection by their nature, and do so either directly or indirectly, in a way different to that in non-gameful actions. Players are more likely to be kept engaged and to keep on playing when their need to belong is fulfilled~\cite{Przybylski2010AEngagement}, as the nature of in-game relationships can impact their behaviors and participation~\cite{Voida2010TheGaming}. Social play can produce feelings of well-being and bring about a performance increases in the players~\cite{Pirker2018AnalyzingDestiny}. Well-designed social game mechanics can result in the players having a strong motivation to complete their tasks and to be retained for longer in the game: they are more motivated to have success~\cite{Nacke2015SocialGame}. Within virtual environments, players can form long-lasting friendships within games, which can continue not only in the real world but also in other games~\cite{Crenshaw2014WhatsGames}.

In this section, we review the literature on \ac{sna} in games to show how researchers examined player networks within specific games. We then focus on the studies analyzing the game provider Steam, and identify a research opportunity in the investigation of the Steam friends network filtered by each of the games studied.

\subsection*{Social Network Analysis in Games}
Researchers and designers can rely on \ac{sna} to monitor the status of in-game social interactions. Player social relationships can be modeled using graphs, which successfully represent interaction patterns among a group of people~\cite{Saltz2004StudentGraphs}. Social networks manifest when directed (e.g., following an account) or indirected (e.g., befrending another user) social interactions are allowed. Social media and standard \ac{osns} explicitly define connections among the users, linked because they are related or share interests. On the other hand, despite being originated from indirect connections, implicit social networks are a rich source of information, and they may thus enforce similar social rules. For instance, online multiplayer games, which are a social phenomenon, encourage social interactions. These interactions, or relationships, can also be interpreted as social networks, thus being modeled using traditional \ac{sna} techniques. Online multiplayer games convey information on the social aspect of gaming~\cite{Freeman2016MakingCommunity} and help understanding social relationships in a highly digitalized world~\cite{Ducheneaut2007VirtualGames}. The study of how players socialize through games can lead to better social environments in games~\cite{Ducheneaut2007VirtualGames,Ducheneaut2004TheGame}. Studying the player network in the form of a graph for example, can highlight how players' activity is reflected in the experiences of others~\cite{Xu2011SociableGame} and how the permanence of certain players can condition others' behaviors~\cite{Xu2011SociableGame}. An inspection of the player network hinted that spending more time in teams is not a synonym of being more social, since the interests of players in social interactions may be purely functional to the game~\cite{Huang2013FunctionalGames,Ducheneaut2004TheGame}. Similarly, toxic interaction patterns may emerge from the analysis~\cite{Jiang2016WhyGames}. Although the employment of SNA in games is still young, researchers have already analyzed the social roles of players. Group formation represents not only a pillar of the player community~\cite{Ducheneaut2007VirtualGames} but sometimes loyalty to the guild also led players to prioritize its growth over their own personal interests~\cite{Ang2010SocialGuilds}. The team organization and connectedness also benefit individual performance~\cite{Kim2017WhatLegends} and retention~\cite{Pirker2018AnalyzingDestiny}. Multiplayer, or social, games foster social relationships by their nature and can thus also be modeled in a graph. Researchers studied groups and community~\cite{Ducheneaut2006AloneGames}, investigating the impact of social structures in gameplay~\cite{Park2015SocialGame,Pirker2018AnalyzingDestiny}. Properties of groups and guilds, for example, are indicators of players' in-game activity and retention~\cite{Rattinger2016IntegratingDestiny,Schiller2018InsideActivity}. Although the player communities formed around games comprise important information about the social dynamics that occur~\cite{Wallner2019BeyondWebsite}, they say much concerning the network composition of the game under investigation. 

\subsection*{Data Analysis on Steam}
Games are multifaceted and as such they engage and connect users in different ways. Consequently, analyzing interaction patterns within games provides a context-specific view of players' social dynamics. Conversely, investigating a game provider could give us a higher-level perspective of players' relationships.

Steam is an online distribution platform for video games. Games on Steam are usually uploaded by the game developers, game studios, or publishers. Each game is presented on a store page. This page contains information about the game such as screenshots, descriptions, developer and publisher information, release year, genre, tags, a list of user reviews, a user rating obtained from the user reviews. % Users can give either a positive or a negative review; the ratings are calculated as a ratio of these ratings. A game's rating is described on a scale from overwhelmingly negative (0\%-19\% of positive reviews) to overwhelmingly positive (95\%-100\% of positive reviews). Each user has a profile page. This page contains information about the user, including friendship information and games played. By default, the information is public. Users can change their profiles to private to hide the information from the public. Users can create and join groups (e.g., to talk about specific games or share information about genres). On Steam, games are distributed, and players can download them and play them directly through the platform. 
Steam also enables connecting to other players. We can gain various insights about players and games by looking at this heterogeneous dataset. Several research teams have used Steam data to understand player preferences~\cite{Zuo2018} and play patterns~\cite{Sifa2015Large-ScaleSteam}. %For instance, the analysis of games reviews conveyed knowledge on players' experiences and expectations towards games or entire genres~\cite{sobkowicz2016steamreviews,Baumann2018}, and indicates that reviews are an excellent source to learn more about games (e.g., design faults)~\cite{Lin,Zuo2018}. Besides the possibility of retrieving players' explicit feedback (reviews), their activity and social relationships can also be studied. 
For instance, clustering playtime allowed the identification of connections among game genres and how players span across different genres~\cite{Sifa2015Large-ScaleSteam}. The social structure of the Steam player network has also been analyzed over the years. The Steam friends network was a modest loosely connected graph in 2011~\cite{Becker2012AnEvolution} and grew substantially in 2016~\cite{ONeill2016}, in the course of which players were found to befriend other players who they found similar to themselves and who also favored social games. This indicates that players are indeed interested in social interactions in games and game platforms. In the context of Steam, researchers mostly studied the activities and behaviors of players to achieve a greater knowledge of the individuals. %Thus, those studies are mostly finalized to improve players' experience by providing ad hoc suggestions or the extraction of interaction patterns within the system. %Little is known on existing social dynamics, and more in-depth studies are needed to grasp how players connect and relate to one another entirely. For instance, researching social contagion and influence within the Steam player network could lead to individuals able to maintain the network alive and connected, resulting in a more profound virtual social connection, serving people's desire to connect. 
However, little is known on how players engaging in the same game are connected and the relationship among those sub-communities, and the specific properties of the games.

\section{Data} %\section{Materials and Method}
In this section, we detail the construction of the data set used to perform our study. Starting by describing how we chose the Steam users and how we built their friendship network, we then detail how we picked the games we considered in this analysis and their \textit{induced} friends' network.

\subsection{The Steam Friendship Graph}
We modeled the Steam friends network as an undirected, non-weighted graph, in which the nodes are the players and the edges represent a friendship status between the two nodes. In the following, we detail the data set construction process. Our intention was to increase the likelihood of the presence of active users. The rationale behind this choice is that regular players actually build a community around a given game or a set of games. On the other hand, the inclusion of inactive players in the friendship network would only modify its structure without contributing to the underlying social dynamics. 
%; for instance, a user may buy a game....

\begin{itemize}
    \item 
        \textit{Step 1.} %To increase the likelihood of the presence of active users within the data set,
        We randomly collected a seed set of $1k$ users from the authors of the reviews of the top $100$ \textit{``New and Trending"} games on the 10th of April 2020. %We did not sample random users as empirical studies discussed how the majority of the users of a social network are passive~\cite{Gong2015,Yang2018}.
        We collected the seed users' friends and built a first friendship network, resulting in a sample of about $50k$ nodes.
        
    \item 
        %To evaluate whether and to what extent a given user influence another player, we needed to build a friendship network not too sparse. For this reason, 
        \textit{Step 2.} We then extracted the largest connected component (LCC) of the first friendship network, consisting of $2.8k$ users. 
        
    \item 
        \textit{Step 3.} We further retrieved the friends of the $2.8k$ users in the LCC, obtaining a second friendship network with $240k$ nodes.
        
    \item 
        \textit{Step 4.} For the last iteration, we also retrieved the friends of the newly added nodes. However, we did not store any new node (i.e., player) in the graph; we only integrated the network with the missing edges. Finally, we removed the users with private profiles.
\end{itemize}
The network obtained by this means counted ${191,479}$ nodes and ${1,242,093}$ edges. We then collected daily updated information about the activities of each player(node) in the network and did this in the form of time spent playing each game they owned. The observation period covered five weeks, from April 13th 2020 to May 17th 2020. Given that we consider users to be active if they have played at least one game during the period of observation, we found that only $51k$ out of $191k$ players were active during the five weeks we crawled their activity. We removed all inactive users from the friendship graph, along with the nodes that became disconnected. This resulted in a final friendship network of $\textbf{39,354}$ users and $\textbf{218,432}$ edges. 
%
%\checkcontent{do we need to add other basic stats? }

\subsection{Games' Friendship Graphs}
% 1. how many games: how did we select them
% 2. induced friendship networks
% 3. basic stats on these networks
We analyzed the friendship network induced by each Steam game to study whether game characteristics, such as genres and tags, reflect on friendship ties of its players.
We picked the $200$ most frequently played games within our dataset and for each of these games we consider a subgraph of the friendship network comprised only the users who have played that game. Thus, we have an edge between two players in each game subgraph if they have both played the given game. We considered only the top $200$ most-played games to ensure a game-induced network of at least $250$ nodes. Table~\ref{tab:game_networks} summarizes the size distribution of the $200$ game networks in terms of the number of nodes (players) and edges (friendship ties).
%
%\checkcontent{do we want other stats here or any comment on the size? }

\begin{table}[t]
\centering
\caption{Distribution of the size of the friendship graphs.}
\label{tab:game_networks}
\begin{tabular}{@{}lccccccc@{}}
\toprule
\multicolumn{1}{c}{} & \textbf{Min} & \textbf{25\%} & \textbf{50\%} & \textbf{75\%} & \textbf{Max} &  \textbf{Mean} & \textbf{Std} \\ \midrule
\textit{\#nodes}  &  $270$ & $369$ & $505$ & $833$ & $17,273$ & $879.79$ & $1,445.36$     \\
\textit{\#edges}  & $7$ &  $38$ & $91$ & $217$ & $32,507$  &  $478.60$ & $2,413.29$ \\
\bottomrule
\end{tabular}

\end{table}

\section{Method}
This section describes the method we followed to detect any pattern in the friendship network induced by specific game characteristics, detailing how we grouped similar networks and the metrics to characterize each group.

\subsection{Graph Clustering}
Our analysis's first step was to group all networks with similar structural characteristics to evaluate whether different games with analogous peculiarities generate similar linking patterns in their induced friendship network. Several clustering algorithms have been proposed in the literature to accomplish this task (i.e., grouping items). However, they all deal with data described by numerical vectors. 

One way to describe graph-structured data through a numerical representation is by engineering handcrafted features. The key characteristics of the networks are thus manually chosen and their values will compose the corresponding numerical vector. Although it is straightforward, this approach may introduce a bias towards the selected features and propagate it in the results. Further, latent patterns encoded in the network may be overlooked or missed.
For this reason, we opted for a graph embedding technique to transform a friendship network in its numeric representation~\cite{Cai_TKDE_2018}. Given a graph $G=(V,E)$ and a predefined embedding dimension $d$, with $d \ll |{V}|$, the problem of graph embedding is to map $G$ into a $d$-dimensional space (a.k.a. latent space), in which the structural properties of $G$ are preserved as much as possible. Following this definition, each graph is represented as either a $d$-dimensional vector (for a whole graph) or a set of $d$-dimensional vectors. Each vector represents the embedding of part of the graph, such as nodes, edges, or substructures. 

As we were interested in clustering graph structures, we used \texttt{graph2vec}~\cite{Narayanan_MLG_2017}, a neural-network-based technique for whole graph embedding. The underlying idea of this graph representation learning process is to yield similar embeddings for structurally similar graphs. Based on the document embedding model~\cite{Mikolov_ICLR_2013}, graph2vec sees each graph as a document and the rooted subgraphs around every node in the graph as words that compose the document. The graph2vec embeddings are task agnostic and as such can be directly used across all analytics tasks involving whole graphs.

Once we obtained the $200$ vector embeddings - one for each game friendship graph - we fed them to the \texttt{K-means} clustering algorithm. We used the Silhouette analysis~\cite{Rousseeuw_JCAM_1987}, giving information about the separation distance between the resulting clusters and the distortion index, counting the sum of squared distances of samples to their closest cluster center, to select the best number of clusters.

\subsection{Clusters Characterization}\label{subsec:characterization}
%\subsection*{Constructs and Metrics}
As a means of characterizing the properties of each cluster, we defined two sets of features considering i) the structural characteristics of the game friendship networks within each group and ii) the metadata associated with each game. This approach is similar to the approach presented by Overgoor et al.~\cite{Overgoor_ICWSM_2020}, where the authors analyze the impact of U.S. college dynamics over students' Facebook friendship relations.
From the structural point of view, we analyzed the following features:
\begin{itemize}
    \item size - number of nodes;
    \item edge density - the share of node pairs that are connected;
    \item mean and variance of node degree distribution;
    \item degree assortativity - which measures the similarity of connections in the graph with respect to the node degree; %(Newman 2003);
    \item group degree centralization - which equals $1$ in a star graph, $0$ in a circle graph where all degrees are equal; %\todo{def/ref}
    \item group betweenness centralization -  which reaches its maximum value $1$ for the star graph; %\todo{def/ref}
    \item average clustering coefficient - which measures the connectedness of a network counting the number of triangles a node is actually involved in overall possible triangles in its neighborhood;  %(Watts and Strogatz 1998);
    \item number of connected components and size of the LCC;
    \item modularity - which measures how good a given graph partition is based on the number of inter- and intra-community edges;%(Clauset, Newman, and Moore 2004);
    \item percentage of the game networks following a power-law distribution.
\end{itemize}
Regarding the game metadata, we considered:
\begin{itemize}
    \item Steam-defined game \textit{genres};
    \item \textit{user-defined tags}, indicating game mechanics, genres, themes, or attributes. Generally, those tags can be any term or phrase\footnote{https://store.steampowered.com/tag/}. As we had an excessive number of tags for each game ($mean=38.57$, and $std=30.22$), we used Term Frequency Inverse Document Frequency~\cite{Manning_2008} (TD-IDF) to obtain a more representative list of user tags for each game. TD-IDF is a numerical statistic intended to reflect how important a word is to a document in a collection or corpus. Hence, we used this metric to retrieve the most characterizing tags (words) for each game (document) in our game dataset (corpus).

    %To obtain a more representative list of user tags for each game, we used Term Frequency Inverse Document Frequency~\cite{Manning_2008} (TD-IDF) to retrieve the most characterizing tags (words) for each game (document).
\end{itemize}
%\checkcontent{check features' description. do we need to further describe why we used TF-IDF?}

%(cleaning the user tags (tfidf) - how, what and why)

%
%
%
\section{Results}

\subsection{Games' Friendship Graphs Clustering}
\begin{figure}[]
    \centering
    \includegraphics[width=0.43\textwidth]{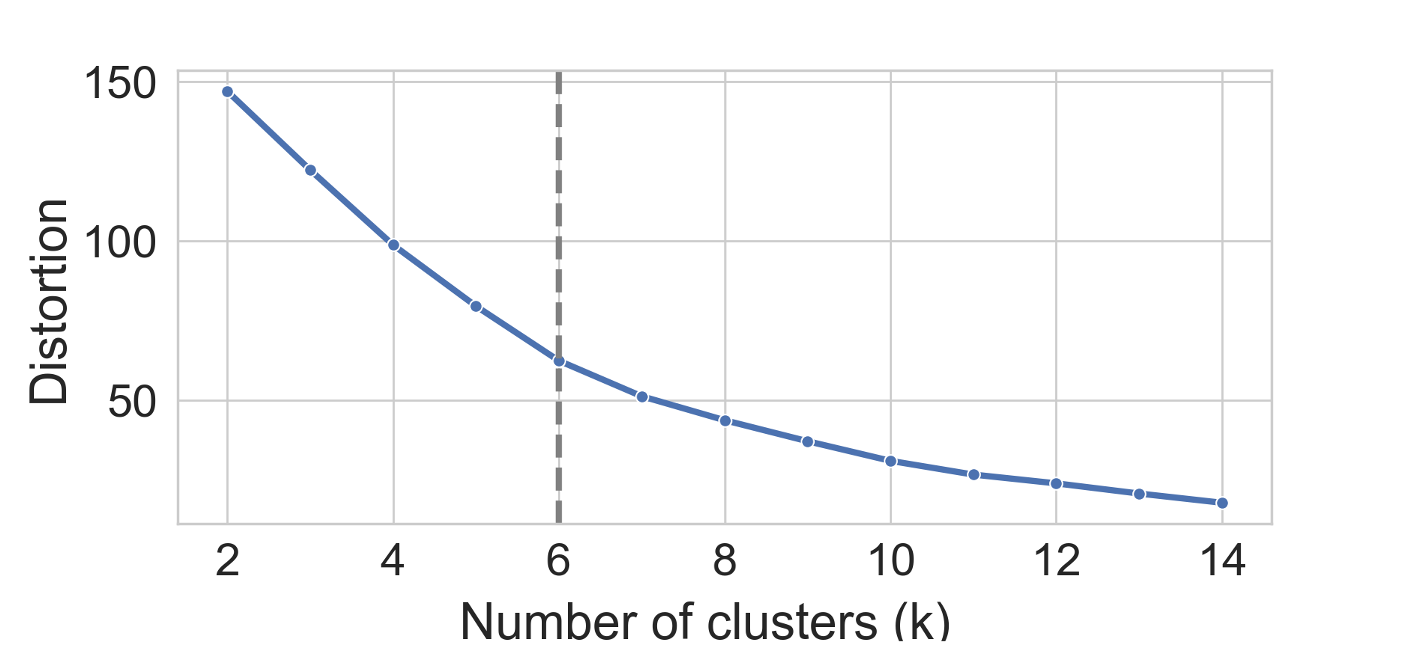}
    \caption{The Elbow Method showing the optimal k.}
    \label{fig:elbow}
\end{figure}

To obtain the network embeddings, we use the implementation of graph2vec at the following link \footnote{https://github.com/benedekrozemberczki/graph2vec}. We run the script using the default parameters suggested by the authors to get an $8$-dimension graph embedding. Regarding the execution of the clustering algorithm, we use the \texttt{sklearn} Python implementation of K-means\footnote{https://scikit-learn.org/stable/modules/generated/sklearn.cluster.KMeans.html}. We iterated the algorithm initializing $k$ from $2$ to $10$. Figure~\ref{fig:elbow} shows the distortion index associated to each value of $k$. The elbow does not clearly denote the best $k$ (either 6, 7, or 8 are valid choices). We decided to set $k=6$, since increasing $k$ led to a higher number of clusters of size~$1$.

\begin{table*}[]
\centering
\caption{Network basic statistics, computed as the average value of each network graph in the cluster.}
\label{tab:net_stats}
\begin{tabular}{@{}lrrrrrrrrrrrr@{}}
\toprule
 \multicolumn{1}{c}{\textbf{Cluster}}&
  \multicolumn{1}{c}{\textbf{nodes}} &
  \multicolumn{1}{c}{\textbf{density}} &
  \multicolumn{1}{c}{\textbf{mean deg}} &
  \multicolumn{1}{c}{\textbf{std deg}} &
  \multicolumn{1}{c}{\textbf{avg clust}} &
  \multicolumn{1}{c}{\textbf{\#CC}} &
  \multicolumn{1}{c}{\textbf{\%LCC}} &
  \multicolumn{1}{c}{\textbf{modularity}} &
  \multicolumn{1}{c}{\textbf{assortativity}} &
  \multicolumn{1}{c}{\textbf{\%pl}} &
  \multicolumn{1}{c}{\textbf{deg centr}} &
  \multicolumn{1}{c}{\textbf{betw centr}} \\ \midrule
\textit{\#0} (106) & 742  & 0,0014 & 60,95 & 145,10 & 0,030 & 583,92  & 0,129 & 0,727 & 0,110 & 63\%  & 5,02E-05 & 0,016 \\
\textit{\#1} (1) & 5.299  & 0.0005 & 17,78 & 147,21 & 0,067 & 2260    & 0,495 & 0,637 & -0,06 & 100\% & 1,05E-05 & 0,171 \\
\textit{\#2} (72) & 395   & 0,0005 & 84,58 & 132,22 & 0,009 & 363,04  & 0,021 & 0,863 & 0,169 & 88\%  & 3,06E-05 & 0,001 \\
\textit{\#3} (13) & 2.741  & 0,0004 & 74,49 & 271,93 & 0,042 & 1773,78 & 0,179 & 0,769 & 0,093 & 62\%  & 6,76E-06 & 0,019 \\
\textit{\#4} (1) & 5.419  & 0,0004 & 27,23 & 214,70 & 0,052 & 2964    & 0,325 & 0,481 & 0,017 & 0\%   & 6,68E-06 & 0,068 \\
\textit{\#5} (1) & 17.273 & 0,0002 & 48,65 & 390,76 & 0,107 & 3783    & 0,711 & 0,617 & 0,062 & 100\% & 1,17E-06 & 0,059 \\ \bottomrule
\end{tabular}
\end{table*}

\begin{figure*}
     \centering
    \begin{subfigure}[t]{0.25\textwidth}
        \raisebox{-\height}{\includegraphics[width=\textwidth]{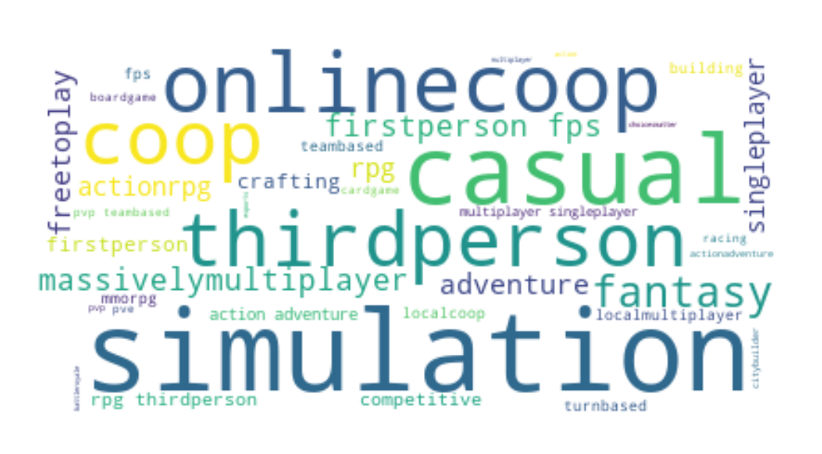}}
        \caption{Cluster 0 (106 games)}
    \end{subfigure}
    \hfill
    \begin{subfigure}[t]{0.25\textwidth}
        \raisebox{-\height}{\includegraphics[width=\textwidth]{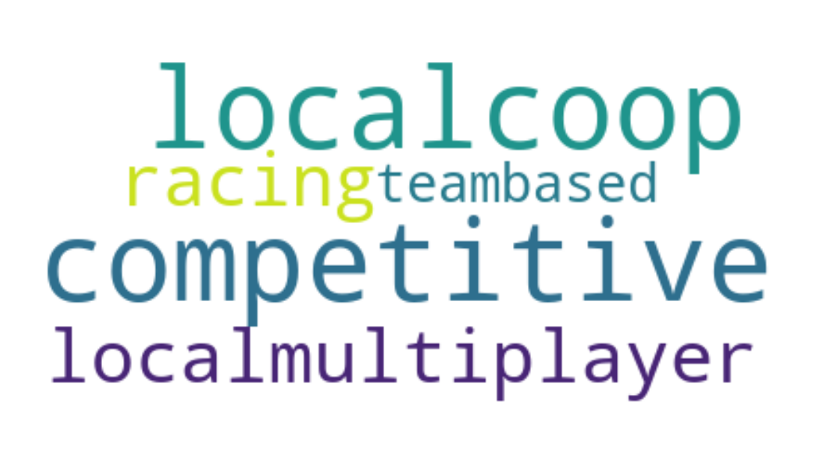}}
        \caption{Cluster 1 (Rocket League)}
    \end{subfigure}
    \hfill
    \begin{subfigure}[t]{0.25\textwidth}
        \raisebox{-\height}{\includegraphics[width=\textwidth]{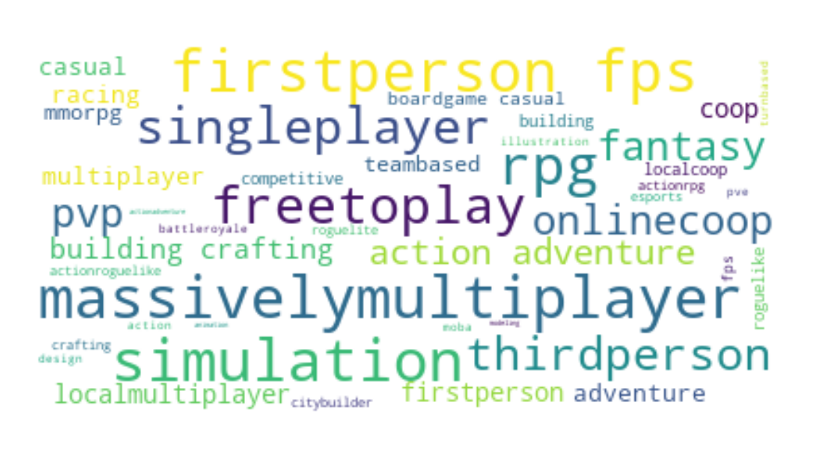}}
        \caption{Cluster 2 (72 games)}
    \end{subfigure}
    %%%%%%%%%%%%%%%%%%%%%%%%%%%%%%%%%%%%second row
    \begin{subfigure}[t]{0.25\textwidth}
        \raisebox{-\height}{\includegraphics[width=\textwidth]{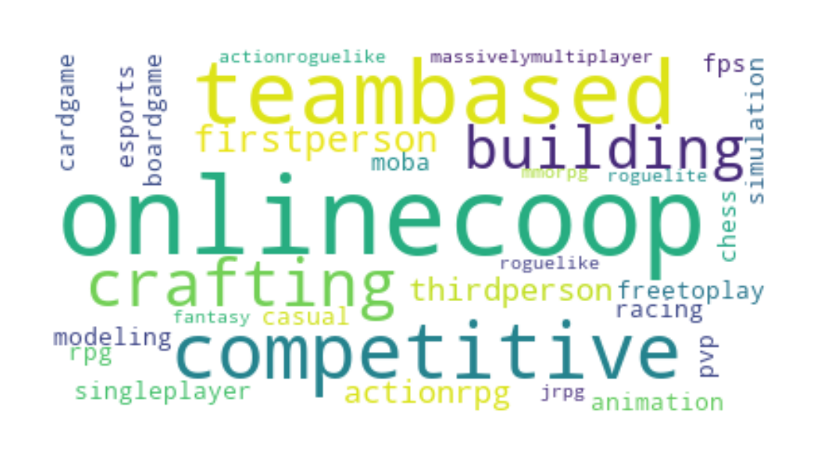}}
        \caption{Cluster 3 (13 games)}
    \end{subfigure}
    \hfill
    \begin{subfigure}[t]{0.25\textwidth}
        \raisebox{-\height}{\includegraphics[width=\textwidth]{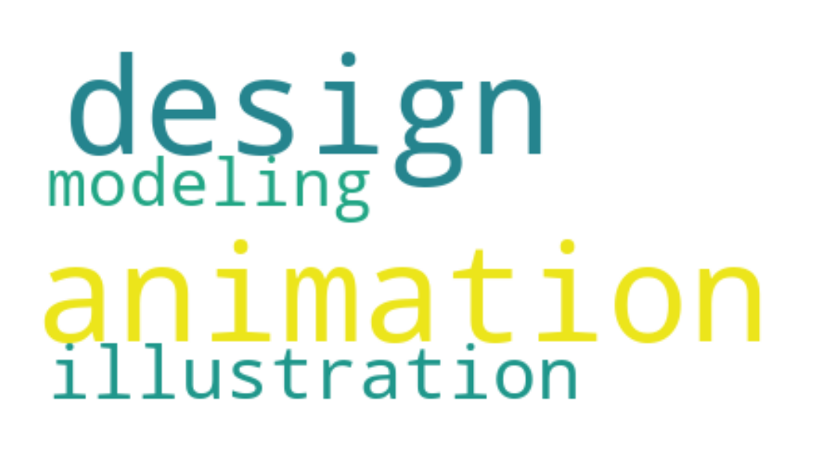}}
        \caption{Cluster 4 (Wallpaper Engine)}
    \end{subfigure}
    \hfill
    \begin{subfigure}[t]{0.25\textwidth}
        \raisebox{-\height}{\includegraphics[width=\textwidth]{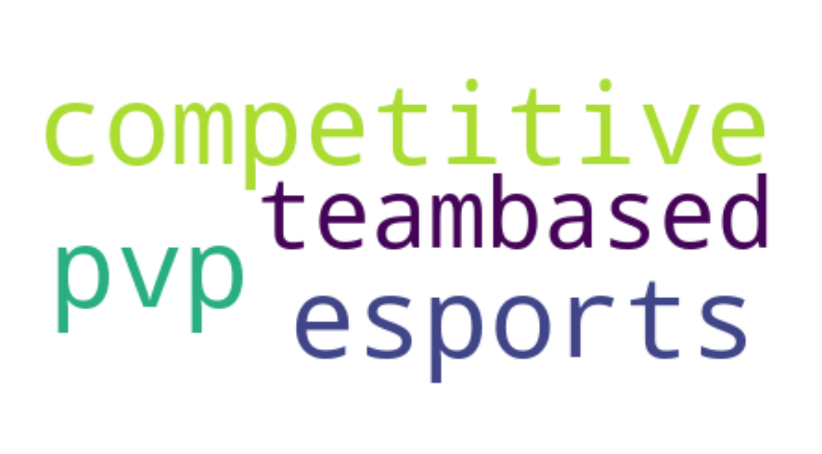}}
        \caption{Cluster 5 (CS:GO)}
    \end{subfigure}
    \caption{Word clouds for the 5 game clusters, displaying the frequency of the user-defined tags in Steam.}
    \label{fig:clouds}
\end{figure*}

\subsection{Characterizing the Game Clusters}
Having clustered the games into six clusters (or groups), we then analyzed the games' properties to characterize each group (see Section~\ref{subsec:characterization}). The structural properties of each cluster are summarized in Table~\ref{tab:net_stats}.

\begin{itemize}
    \item \textbf{Cluster 0.} In this cluster we have 106 games, which show the most \textit{dense} networks, but with a \textit{low clustering} coefficient. The network is comprised of many connected components and the largest of these covers about 10\% of the nodes. This division in many components is also reflected by the \textit{modularity} score, which is among the highest. The games in the cluster present a \textit{centralized structure}, in terms of degree. The nodes, on average, present a high degree distribution but very variable ($mean = 60$, and $std = 145$).
    \item \textbf{Cluster 1.} This cluster includes one game: Rocket League. The network is made of about $5$k nodes, is not particularly dense, and has the \textit{lowest degree distribution} ($mean = 17$ and $std = 147$). The degree distribution follows the power-law distribution, and thus the network is \textit{scale-free}. Although many communities exist in this network, the largest connected component accounts for half of the nodes. Finally, the network is \textit{centralized}, in terms of betweenness centrality.
    \item \textbf{Cluster 2.} In this cluster, we have 72 games, which have (on average) the smallest network size. These networks have the \textit{highest mean degree} distribution ($mean = 84$, $std = 132$), but the lowest average clustering coefficient. The networks are \textit{scattered} in many separate connected components, with the biggest covering 2\% of the network, and thus show the highest \textit{modularity} score. The game networks in this cluster also have the highest degree assortativity, and most of them \textit{(88\%) are scale-free} networks. Finally, they are relatively centralized, although not as much as the networks in Cluster 0.
    \item \textbf{Cluster 3.} In this cluster, we have 13 games. The networks are, on average, much \textit{bigger} than in Cluster 0 and 2. The degree distribution has a quite high mean, although being very variable ($mean = 74$, and $std = 271$). The networks show a high \textit{modularity} and are divided into\textit{ many components}, with the largest accounting for 18\% of the nodes. Finally, the networks are not particularly centralized, and only about half of them are scale-free.
    \item \textbf{Cluster 4.} This cluster includes one title: Wallpaper Engine, whose network counts about \textit{5k nodes}. The \textit{degree distribution is quite low}, but very variable ($mean = 27$, and $std = 214$), and does not follow the power-law distribution. The network shows the lowest modularity score and is not particularly centralized.
    \item \textbf{Cluster 5.} This cluster includes only Counter-Strike: Global Offense (CS:GO), the network of which is much \textit{bigger} than the others (17k nodes). The network has the highest clustering coefficient and a \textit{big LCC} (71\% of the network). The CS:GO network is also a \textit{scale-free} network, with the \textit{lowest centralization} score.
\end{itemize}

\begin{table*}[h!]
\centering
\caption{Summary of the clusters and their characterization --- the percentages in brackets indicate how frequent the game type was in the cluster, for the clusters with more than one element.}
\label{tab:summary_clusters}
\begin{tabular}{@{}ccll@{}}
\toprule
\multicolumn{1}{l}{\textbf{Cluster}} &
  \multicolumn{1}{l}{\textbf{Size}} &
  \textbf{Game type (from user-defined tags)} &
  \textbf{Game network} \\ \midrule
\textit{\#0} &
  106 &
  Simulation (100\%), online coop (70\%), casual (70\%) &
  Dense, low clustering coeff, centralized structure (degree) \\ 
\textit{\#1} &
  1 &
  \begin{tabular}[c]{@{}l@{}}Local coop, local multiplayer, team-based (RL)\end{tabular} &
  Scale-free, low degree, centralized structure (betweenneess), LCC 50\% \\ 
\textit{\#2} &
  72 &
  \begin{tabular}[c]{@{}l@{}} Massively multiplayer (100\%), F2P (90\%), \\singleplayer (73\%) \end{tabular}  &
  88\% scale-free, scattered, high modularity, low clustering coeff, small networks \\
\textit{\#3} &
  13 &
  Online coop (100\%), team-based (70\%) & %, competitive (50\%)
  High modularity, LCC 18\%, quite big networks \\
\textit{\#4} &
  1 &
  \begin{tabular}[c]{@{}l@{}}Not a game but a design tool (Wallpaper Engine)\end{tabular} &
  5k players, but low degree, lowest modularity \\
\textit{\#5} &
  1 &
  \begin{tabular}[c]{@{}l@{}}Esport, team-based, competitive (CS:GO)\end{tabular} &
  17k players, scale-free, highest clustering coeff, LCC 71\%, not centralized \\ \bottomrule
\end{tabular}
\end{table*}

In addition to inspecting the network features, we also studied how the tags defined by the users in Steam are distributed in each cluster. In order to do this, we build a word cloud for each cluster (Figure~\ref{fig:clouds}). 
The clouds are obtained through the \texttt{wordcloud} python package, and compute the word (i.e., tag) frequency in the text (i.e., cluster tag list). Cluster 1, 4, and 5 include only one game each, and thus the tag displayed entirely describe that specific game title. Rocket League (\texttt{Cluster 1}) is a team-based, \textit{competitive} game, which often is used with the \textit{local multiplayer} feature to build the teams. Wallpaper Engine (\texttt{Cluster 4}) is a \textit{modeling} and design tool to make animated wallpaper for windows. Finally, CS:GO (\texttt{Cluster 5}) is a team-based \textit{competitive} game, which falls under the category of \textit{esport}. On the other hand, the other word clouds are more crowded. \texttt{Cluster 0} has mostly \textit{cooperative}, \textit{casual} games, which are in third person and/or simulation. In \texttt{Cluster 2}, the differences are less delineated. Nevertheless, the games are mostly free-to-play, with a \textit{massively multiplayer} and \textit{single-player} mechanic combined. Finally, \texttt{Cluster 3} is characterized by \textit{team-based} games, both \textit{cooperative} and \textit{competitive}, with elements of crafting and building.

The game genres are uniformly distributed across the clusters. Except for Cluster 1, 4, and 5, whose genres reflect the single game that they represent, the other clusters lack a narrow distinction of the characterizing genres. Specifically, the most popular genre was \textit{Action} (17\% for Cluster 0, 13\% for Cluster 2, and 30\% for Cluster 3), followed by \textit{Adventure} (13\% for Cluster 0, 12\% for Cluster 2, and 9\% for Cluster 3), \textit{Simulation} (12\% for Cluster 0, 10\% for Cluster 2, and 15\% for Cluster 3), and \textit{RPG} (13\% for Cluster 0, 8\% for Cluster 2, and 8\% for Cluster 3). The genre property thus does not contribute to the cluster characterization as much as user-defined tags did.
Table~\ref{tab:summary_clusters} summarizes these findings.

\section{Discussion}
In this work, we analyzed the Steam friendship graph for 200 games to connect their network properties to their characteristics (i.e., genre and user-defined tags). Our investigation led to six clusters, summarized in Table~\ref{tab:summary_clusters}.

We observed how user-defined tags were more informative than game genres, as the genres are distributed in similar ways across the six clusters. Although we should take into consideration the fact that the user-defined tags also go down to a greater level of detail, this result suggests how the way players play games is related to the structure of the community. The game network analyses can thus provide valuable insights into how the game is perceived and interpreted by its community and its dominant social dynamics. In other words, the players are essential for shaping the gaming experience and its mechanics on a level similar to that of the designers. This finding also emphasized the importance of analyzing the player network not only to detect interaction patterns, but also to some extent for obtaining design feedback. 

The cluster characterizations also provided more in-depth correlations among the community structures and the game properties. First, we observed that team-based games tended to have scale-free networks, regardless of the network sizes (Cluster 1 and 5). In other words, the degree distributions of their nodes follow a power-law distribution, in which many nodes have few connections but only a few nodes have a large number of links. This is also known as the richer-get-richer phenomenon, prevalent in social media networks. Therefore, the network structure of games with a strong team component shows some similarities to social media communities~\cite{mislove2007measurement,Barabasi_2016}, suggesting that playing in teams creates an environment in which social relationships are nurtured. The existence of teams in itself also impacted the clustering coefficient, which was higher than in cooperative but casual games (Cluster 0). The result here is that playing in teams increases the likelihood of becoming friends of friends in Steam than sharing a similar game preference (either cooperative or single-player). Additionally, multiplayer games without a strong team-based component (Cluster 0 and 2) tended to have much more centralized networks than games in which teams are important (Cluster 1, 3, and 5). Hence, having players' connected in teams uniformed the link distribution, increasing the likelihood of forming triangles (i.e., friends of friends). 

Most of the massively multiplayer games with a single-player component (Cluster 2) also showed a tendentially scale-free network structure. These networks were highly scattered in small graphs with high modularity, however, whereas team-based game graphs were more connected. This finding further strengthens the importance of teams or team tasks to obtain a more cohesive community, especially in terms of relationships beyond the game (i.e., a Steam friendship). Furthermore, the scattered nature of these networks can be linked to them being massively multiplayer games. The result of this is that players come into contact with many (random) users. Consequently, it is unlikely that they will form a (Steam) friendship or play in more matches together. The graph of Cluster 4, which represents a design tool only usable in a single-user mode, is similarly scattered and sparse. This characterization lends support to the importance of an in-game social aspect for nurturing the game community. 

Finally, the results show that the biggest network with the highest clustering coefficient is a team-based esport game (CS:GO). Games of this type rely on the idea of (consolidated) teams more than others and do so not only for professionals but also for amateurs. This might contribute to the high clustering coefficient, as repeatedly playing together might result in a Steam friendship (or vice-versa friends form stable teams).

In summary, the shape of the game community is more affected by how players define the game (through user-defined tags) than by the game genre assigned by the designers/producers. Specifically, games that require the formation of teams show specific network properties: they are not centralized, have a high clustering coefficient, and tend to be scale-free. They support the formation of friendships to a greater extent than non-team-based games. Conversely, titles with a stronger single-player focus have more scattered networks, made of many small detached components.

\subsection*{Limitations}

This study also comes with a few limitations. First, our population sample comprehends a subset of the actual Steam network. Consequently, we have a partial view of the friends graph and the network in each game. Additionally, our sample is seeded from the players reviewing ``new and trending games'' and playing at least one game within our observation period. This means we must interpret our results in terms of active players (in our 5-week time window). Furthermore, the resulting friendship networks may also have been partially influenced by the recent pandemic since more people may have been obliged to stay at home and will have had more time to play games during this period.
Second, the game networks vary in size and the sub-graph structure might not entirely reflect the full-game network, in particular for the smaller networks. Since some network-related features could have accentuated this bias, we relied on graph embeddings to perform the comparison. Finally, we do not consider the temporal component in our study. Hence, we cannot infer causality in the relationships between graph topology and friendship connections. In other words, we can neither argue that players became friends because they played the same game title or that they began playing because they were already friends. Rather, we correlate the Steam community structure of each game to the game properties to identify patterns linking the network properties and game type (i.e., user tags). 

\subsection*{Future Works}
The natural next step for our study will be the introduction of the temporal factor in the analysis. In this study, we cannot infer causality among the two variables studied: the game network structure and the game tags. In future work, we aim to achieve an understanding of the link between game tags and how they can be used to predict new games, and vice-versa (i.e., the link between game tags and how they can be used to predict new friends). Additionally, we will analyze the network evolution for each game to understand the dynamics by which the connections are formed (e.g., velocity and patterns). Finally, we will investigate how we can infer the (used) game mechanics and playstyle from the network structure. In this regard, we will verify how users play the game despite the available mechanics. For example, a game might be both cooperative and competitive, but the network might prefer building a community through cooperation.

\section{Conclusion}
People connect through games, in which they build relationships and friendships that can be translated into other games and also into real world situations. Recent research on the player networks induced from in-game social interactions showed the potential of \ac{sna} in games towards unraveling the dynamic occurring within the game community. Nevertheless, a broader understanding of how those game networks are shaped and formed still lacks. Also lacking is an analysis of how the game characteristics are related to the structure of its player network. This study deepened our understanding of player communities by analyzing the Steam friendship graphs for 200 games. We found that some game characteristics are linked to specific network properties. The game community's shape is related to the tags users associated with games, more than the game official genres. Specifically, team-based games have scale-free networks that are not centralized and with a high clustering coefficients. Hence, they support the formation of friendships to a far greater extent than non-team-based games. On the other hand, single-player games have scattered networks made of many small detached components.
In conclusion, the type of game is indeed reflected in how the community is shaped and formed, supporting the idea that playing together (especially in teams) is more likely to produce social capital, and thus (Steam) friendships.

\bibliographystyle{IEEEtran}
\bibliography{main}

% Generated by IEEEtran.bst, version: 1.14 (2015/08/26)
\begin{thebibliography}{10}
\providecommand{\url}[1]{#1}
\csname url@samestyle\endcsname
\providecommand{\newblock}{\relax}
\providecommand{\bibinfo}[2]{#2}
\providecommand{\BIBentrySTDinterwordspacing}{\spaceskip=0pt\relax}
\providecommand{\BIBentryALTinterwordstretchfactor}{4}
\providecommand{\BIBentryALTinterwordspacing}{\spaceskip=\fontdimen2\font plus
\BIBentryALTinterwordstretchfactor\fontdimen3\font minus
  \fontdimen4\font\relax}
\providecommand{\BIBforeignlanguage}[2]{{%
\expandafter\ifx\csname l@#1\endcsname\relax
\typeout{** WARNING: IEEEtran.bst: No hyphenation pattern has been}%
\typeout{** loaded for the language `#1'. Using the pattern for}%
\typeout{** the default language instead.}%
\else
\language=\csname l@#1\endcsname
\fi
#2}}
\providecommand{\BIBdecl}{\relax}
\BIBdecl

\bibitem{Johannes2020VideoWell-being}
N.~Johannes, M.~Vuorre, and A.~K. Przybylski, ``{Video game play is positively
  correlated with well-being},'' \emph{Royal Society Open Science}, vol.~8,
  no.~2, p. 202049, 2021.

\bibitem{Przybylski2010AEngagement}
A.~K. Przybylski, C.~S. Rigby, and R.~M. Ryan, ``{A Motivational Model of Video
  Game Engagement},'' \emph{Review of General Psychology}, vol.~14, no.~2, pp.
  154--166, 2010.

\bibitem{Pirker2018AnalyzingDestiny}
J.~Pirker, A.~Rattinger, A.~Drachen, and R.~Sifa, ``{Analyzing player networks
  in Destiny},'' \emph{Entertainment Computing}, vol.~25, pp. 71--83, 2018.

\bibitem{Schiller2018InsideActivity}
M.~H. Schiller, G.~Wallner, C.~Schinnerl, A.~M. Calvo, J.~Pirker, R.~Sifa, and
  A.~Drachen, ``{Inside the group: Investigating social structures in player
  groups and their influence on activity},'' \emph{IEEE Transactions on Games},
  vol.~11, no.~4, pp. 416--425, 2018.

\bibitem{Mandryk2020HowWellbeing}
R.~L. Mandryk, J.~Frommel, A.~Armstrong, and D.~Johnson, ``{How Passion for
  Playing World of Warcraft Predicts In-Game Social Capital, Loneliness, and
  Wellbeing},'' \emph{Frontiers in Psychology}, vol.~11, p. 2165, 2020.

\bibitem{Nacke2015SocialGame}
L.~E. Nacke, M.~Klauser, and P.~Prescod, ``{Social Player Analytics in a
  Facebook Health Game},'' in \emph{Proceedings of KHCI}, 2015, pp. 180--187.

\bibitem{wallner2019tweeting}
G.~Wallner, S.~Kriglstein, and A.~Drachen, ``Tweeting your destiny: Profiling
  users in the twitter landscape around an online game,'' in \emph{2019 IEEE
  Conference on Games (CoG)}.\hskip 1em plus 0.5em minus 0.4em\relax IEEE,
  2019, pp. 1--8.

\bibitem{Wallner2019BeyondWebsite}
G.~Wallner, C.~Schinnerl, M.~H. Schiller, A.~Monte~Calvo, J.~Pirker, R.~Sifa,
  and A.~Drachen, ``{Beyond the individual: Understanding social structures of
  an online player matchmaking website},'' \emph{Entertainment Computing},
  vol.~30, p. 100284, 2019.

\bibitem{Kim2017WhatLegends}
Y.~J. Kim, D.~Engel, A.~W. Woolley, J.~Y.~T. Lin, N.~McArthur, and T.~W.
  Malone, ``{What makes a strong team? Using collective intelligence to predict
  team performance in League of Legends},'' in \emph{Proceedings of CSCW},
  2017, pp. 2316--2329.

\bibitem{loria2020influencers}
E.~Loria, J.~Pirker, A.~Drachen, and A.~Marconi, ``Do influencers
  influence?--analyzing players' activity in an online multiplayer game,'' in
  \emph{International Conference on Games (CoG)}.\hskip 1em plus 0.5em minus
  0.4em\relax ACM, 2020.

\bibitem{Baumeister1995TheMotivation}
R.~F. Baumeister and M.~R. Leary, ``{The Need to Belong: Desire for
  Interpersonal Attachments as a Fundamental Human Motivation},''
  \emph{Psychological Bulletin}, vol. 117, no.~3, pp. 497--529, 1995.

\bibitem{Rogers2017TheApproach}
R.~Rogers, ``{The motivational pull of video game feedback, rules, and social
  interaction: Another self-determination theory approach},'' \emph{Computers
  in Human Behavior}, vol.~73, pp. 446--450, 2017.

\bibitem{Voida2010TheGaming}
A.~Voida, S.~Carpendale, and S.~Greenberg, ``{The individual and the group in
  console gaming},'' in \emph{Proceedings of CSCW}, 2010, pp. 371--380.

\bibitem{Crenshaw2014WhatsGames}
N.~Crenshaw and B.~Nardi, ``{What's in a name? Naming practices in online video
  games},'' in \emph{Proceedings of CHI Play}, 2014, pp. 67--76.

\bibitem{Saltz2004StudentGraphs}
J.~S. Saltz, S.~R. Hiltz, and M.~Turoff, ``{Student social graphs},'' in
  \emph{Proceedings of CSCW}.\hskip 1em plus 0.5em minus 0.4em\relax New York,
  New York, USA: ACM (ACM), 2004, p. 596.

\bibitem{Freeman2016MakingCommunity}
G.~Freeman, ``{Making games as collaborative social experiences: Exploring an
  online gaming community},'' in \emph{Proceedings of CSCW}, vol.
  26-February-2016, 2016, pp. 265--268.

\bibitem{Ducheneaut2007VirtualGames}
N.~Ducheneaut, R.~J. Moore, and E.~Nickell, ``{Virtual "third places": A case
  study of sociability in massively multiplayer games},'' \emph{Computer
  Supported Cooperative Work}, vol.~16, no. 1-2, pp. 129--166, 2007.

\bibitem{Ducheneaut2004TheGame}
N.~Ducheneaut and R.~J. Moore, ``{The Social Side of Gaming: A Study of
  Interaction Patterns in a Massively Multiplayer Online Game},'' in
  \emph{Proceedings of CSCW}, 2004.

\bibitem{Xu2011SociableGame}
Y.~Xu, X.~Cao, A.~Sellen, R.~Herbrich, and T.~Graepel, ``{Sociable killers:
  Understanding social relationships in an online first-person shooter game},''
  in \emph{Proceedings of CSCW}, 2011, pp. 197--206.

\bibitem{Huang2013FunctionalGames}
Y.~Huang, W.~Ye, N.~Bennett, and N.~S. Contractor, ``{Functional or social?
  Exploring teams in online games},'' in \emph{Proceedings of CSCW}.\hskip 1em
  plus 0.5em minus 0.4em\relax New York, New York, USA: ACM Press, 2013, pp.
  399--408.

\bibitem{Jiang2016WhyGames}
J.~Jiang and S.~Yarosh, ``{Why do teammates hate me? Cross-cultural tensions
  and social dynamics in online games},'' in \emph{Proceedings of CSCW}, vol.
  26-February-2016, 2016, pp. 301--304.

\bibitem{Ang2010SocialGuilds}
C.~S. Ang and P.~Zaphiris, ``{Social Roles of Players in MMORPG Guilds},''
  \emph{Information, Communication {\&} Society}, vol.~13, no.~4, pp. 592--614,
  2010.

\bibitem{Ducheneaut2006AloneGames}
N.~Ducheneaut, N.~Yee, E.~Nickell, and R.~J. Moore, ``{"Alone Together?"
  Exploring the Social Dynamics of Massively Multiplayer Online Games},'' in
  \emph{Proceedings of CHI}.\hskip 1em plus 0.5em minus 0.4em\relax New York,
  New York, USA: ACM Press, 2006.

\bibitem{Park2015SocialGame}
H.~Park and K.~J. Kim, ``{Social network analysis of high-level players in
  multiplayer online battle arena game},'' in \emph{Lecture Notes in Computer
  Science}, vol. 8852.\hskip 1em plus 0.5em minus 0.4em\relax Springer Verlag,
  2015, pp. 223--226.

\bibitem{Rattinger2016IntegratingDestiny}
A.~Rattinger, G.~Wallner, A.~Drachen, J.~Pirker, and R.~Sifa, ``{Integrating
  and inspecting combined behavioral profiling and social network models in
  Destiny},'' in \emph{Lecture Notes in Computer Science}, vol. 9926
  LNCS.\hskip 1em plus 0.5em minus 0.4em\relax Springer Verlag, 2016, pp.
  77--89.

\bibitem{Zuo2018}
Z.~Zuo, ``{Sentiment Analysis of Steam Review Datasets using Naive Bayes and
  Decision Tree Classifier},'' Tech. Rep., 2018.

\bibitem{Sifa2015Large-ScaleSteam}
R.~Sifa, A.~Drachen, and C.~Bauckhage, ``{Large-Scale Cross-Game Player
  Behavior Analysis on Steam},'' Tech. Rep., 2015.

\bibitem{Becker2012AnEvolution}
R.~Becker, Y.~Chernihov, Y.~Shavitt, and N.~Zilberman, ``{An analysis of the
  Steam community network evolution},'' \emph{2012 IEEE 27th Convention of
  Electrical and Electronics Engineers in Israel, IEEEI 2012}, 2012.

\bibitem{ONeill2016}
M.~O'Neill, E.~Vaziripour, J.~Wu, and D.~Zappala, ``{Condensing steam:
  Distilling the diversity of gamer behavior},'' \emph{Proceedings of the ACM
  SIGCOMM Internet Measurement Conference}, vol. 14-16-Nove, pp. 81--95, 2016.

\bibitem{Cai_TKDE_2018}
H.~Cai, V.~W. Zheng, and K.~Chang, ``A comprehensive survey of graph embedding:
  Problems, techniques, and applications,'' \emph{IEEE Transactions on
  Knowledge \& Data Engineering}, vol.~30, no.~09, pp. 1616--1637, 2018.

\bibitem{Narayanan_MLG_2017}
A.~Narayanan, M.~Chandramohan, R.~Venkatesan, L.~Chen, Y.~Liu, and S.~Jaiswal,
  ``graph2vec: Learning distributed representations of graphs,'' in
  \emph{Proceedings of the 13th International Workshop on Mining and Learning
  with Graphs}, 2017.

\bibitem{Mikolov_ICLR_2013}
T.~Mikolov, K.~Chen, G.~Corrado, and J.~Dean, ``Efficient estimation of word
  representations in vector space,'' in \emph{1st International Conference on
  Learning Representations, Workshop Track Proceedings}, 2013.

\bibitem{Rousseeuw_JCAM_1987}
P.~Rousseeuw, ``Silhouettes: A graphical aid to the interpretation and
  validation of cluster analysis,'' \emph{Journal of Computational and Applied
  Mathematics}, vol.~20, pp. 53--65, 1987.

\bibitem{Overgoor_ICWSM_2020}
J.~Overgoor, B.~State, and L.~A. Adamic, ``The structure of u.s. college
  networks on facebook,'' \emph{Proceedings of the International AAAI
  Conference on Web and Social Media}, vol.~14, no.~1, pp. 499--510, 2020.

\bibitem{Manning_2008}
C.~D. Manning, P.~Raghavan, and H.~Schütze, \emph{Introduction to Information
  Retrieval}.\hskip 1em plus 0.5em minus 0.4em\relax Cambridge University
  Press, 2008.

\bibitem{mislove2007measurement}
A.~Mislove, M.~Marcon, K.~P. Gummadi, P.~Druschel, and B.~Bhattacharjee,
  ``Measurement and analysis of online social networks,'' in \emph{Proceedings
  of the 7th ACM SIGCOMM conference on Internet measurement}, 2007, pp. 29--42.

\bibitem{Barabasi_2016}
A.-L. Barabási, \emph{Network Science}.\hskip 1em plus 0.5em minus 0.4em\relax
  Cambridge University Press, 2016.

\end{thebibliography}

\newpage
%\appendix{}
\appendix{\textit{A. List of the Steam titles contained in each cluster.}}
\begin{table}[h!]
\centering
%\caption{List of the games contained in each cluster.}
%\label{tab:my-table}
\begin{tabularx}{\columnwidth}{|l|X|}
%\begin{tabular}{ll}
\hline
\textbf{Cluster} & \multicolumn{1}{c|}{\textbf{Steam Titles}} \\ \hline
\#0 &
  Counter-Strike, Counter-Strike: Source, Left 4 Dead 2, Portal 2, Sid Meier's Civilization V, Grand Theft Auto IV: The Complete Edition, Total War: Shogun 2,  Red Dead Redemption 2, FinalFantasy XIV Online, Mount \& Blade: Warband, Borderlands 2, The Elder Scrolls V: Skyrim, Tomb Raider, Alien: Isolation, Payday 2, Grim Dawn, DayZ, Insurgency, Ghostrunner, Euro Truck Simulator 2, Wreckfest, Warframe, Company of Heroes 2,Killing Floor 2, War Thunder, Star Wars Jedi: Fallen Order, Europa Universalis IV, Path of Exile, Dying Light, The Forest, Assetto Corsa, Plague Inc: Evolved, The Binding of Isaac: Rebirth, Cities: Skylines, Tesla Effect: A Tex Murphy Adventure, Mount \& Blade II: Bannerlord, Destiny 2, Fallout 76, XCOM 2,  American Truck Simulator, No Man's Sky, Dota Underlords, Stellaris, 100\% Orange Juice, Sid Meier’s Civilization VI,  Brawlhalla, The Witcher 3: Wild Hunt, Unturned, The Elder  Scrolls Online, Call of Duty: Black Ops III, Geometry  Dash, Don't Starve Together, ARK: Survival Evolved,  Black Mesa, Dark Souls III, Human Resource Machine,  Fallout 4, Doom, Scrap Mechanic, Tekken 7, Rise of the Tomb Raider, Squad, Hearts of Iron IV, Tower Unite,  Borderlands 3, Business Tour - Board Game with Online  Multiplayer, Stardew Valley, Factorio, Golf With Your Friends,  VRChat, Friday the 13th: The Game, Pac-Man Championship Edition 2, Pinball FX3, Paladins, UNO, Cyrano Story,  Human: Fall Flat, Minion Masters, The Elder Scrolls V:  Skyrim Special Edition, Manual Samuel - Anniversary Edition,Ashes of the Singularity: Escalation, Redout: Enhanced Edition, Deep Rock Galactic, Warhammer: Vermintide 2, Playerunknown's Battelgrounds, Insurgency: Sandstorm, Assassin's Creed Origins, Total War: Warhammer, Streets of Rage 4, Hunt: Showdown, Mordhau, Slay the Spire, Jurassic World  Evolution, Generation Zero, Shadow of the Tomb Raider:  Definitive Edition, Project Winter, Doom Eternal, Farming  Simulator 19, Assassin's Creed Odyssey, Age of Empires II:  Definitive Edition, Green Hell, HITMAN 2, Pummel Party,  XCOM: Chimera Squad, Resident Evil 2, Resident Evil 3,  Mortal Kombat 11, Halo: The Master Chief Collection  \\ \hline
\#1 &
  Rocket League \\ \hline
\#2 &
  Half-Life, Half-Life 2, Star Wars: Battlefront 2, Call of Duty: World at War, Grand Theft Auto: San Andreas, Fallout: New Vegas, Call of Duty: Black Ops, Realm of the Mad God Exalt, Call of Duty: Black Ops II, Crusader Kings II, SpeedRunners, Starbound, Bloons TD 6 PlanetSide 2, Age of Empires II, Just Cause 3, Deadside, Heroes \& Generals, Prison Architect, Space Engineers, 7 Days to Die, Darkest Dungeon, Subnautica, tModLoader, BeamNG.drive, Metal Gear Solid V: The Phanthom Pain, Warface, RimWorld, Trove, For Honor, DiRT Rally,  Enter the Gungeon, Town of Salem, Dark Souls II:  Scholar of the First Sin, Elite Dangerous, Astroneer,  Blender, Hollow Knight, Project Cars 2, Smite,  CRSED:  F.O.A.D., The Jackbox Party Pack 2, Rising Storm 2: Vietnam,  Blackwake, The Jackbox Party Pack 3, KovaaK 2.0, F12019,  Divinity: Original Sin 2 - Definitive Edition, Conan  Exiles, Drawful 2, Deceit, Northgard, Planet Coaster,  Half-Life: Alyx, Far Cry 5, Dark Souls: Remastered,  Black Desert Online, Dead Cells, The Jackbox Party  Pack 4, House Flipper, Remnant: From the Ashes,  Beat Saber, Soundpad, Raft, Stick Fight: The Game, Dragon Ball FighterZ, DiRT Rally 2.0, SCP: Secret  Laboratory, Aim Lab, Post Scriptum, Albion Online,  Sekiro: Shadows Die Twice - Goty Edition \\ \hline
\#3 &
  Team Fortress 2, Dota 2, Garry's Mod, Terraria, Arma 3,  SteamVR, Rust, Grand Theft Auto V, Tabletop Simulator, Tom Clancy's Rainbow Six Siege, Dead by Daylight, Monster Hunter: World, Risk of Rain 2 \\ \hline
\#4 &
  Wallpaper Engine \\ \hline
\#5 &
  Counter-Strike: Global Offensive \\ \hline
%\end{tabular}
\end{tabularx}
\end{table}

\end{document}